\documentclass[a4paper,fleqn,usenatbib]{mnras}

\usepackage{newtxtext,newtxmath}
\usepackage[T1]{fontenc}
\usepackage{ae,aecompl}

\usepackage{graphicx}	% Including figure files
\usepackage{amsmath}	% Advanced maths commands
\usepackage{amssymb}	% Extra maths symbols

\title[No AGN evidence in NGC\,1614 from deep VLBI radio observations]{No AGN evidence in NGC\,1614 from deep radio VLBI observations}

\author[Herrero-Illana et al.]{Rub\'en Herrero-Illana,$^{1}$\thanks{E-mail:
rherrero@eso.org}
Antxon Alberdi,$^{2}$
Miguel \'Angel P\'erez-Torres,$^{2,3}$
\newauthor Almudena Alonso-Herrero,$^{4}$
Daniel Gonz\'alez-Mill\'an,$^{2}$ and
Miguel Pereira-Santaella$^{5}$
\\
$^{1}$European Southern Observatory (ESO), Alonso de C\'ordova 3107, Vitacura, Casilla 19001, Santiago de Chile, Chile\\
$^{2}$Instituto de Astrof\'isica de Andaluc\'ia (IAA-CSIC). Glorieta de la Astronom\'ia s/n, 18008, Granada, Spain\\
$^{3}$Visiting Scientist: Departamento de F\'isica Te\'orica, Facultad de Ciencias, Universidad de Zaragoza, Spain\\
$^{4}$Centro de Astrobiolog\'ia, CSIC-INTA,ESAC Campus, E-28692 Villanueva de la Ca\~nada, Madrid, Spain\\
$^{5}$Department of Physics, University of Oxford, Keble Road, Oxford OX1 3RH, UK%\\
}

\pubyear{2017}

\begin{document}
\label{firstpage}
\pagerange{\pageref{firstpage}--\pageref{lastpage}}
\maketitle

% Abstract of the paper    % Max 200 words
\begin{abstract}
We present deep dual-band 5.0 and 8.4\,GHz European VLBI Network (EVN) observations of NGC\,1614, a local luminous infrared galaxy with a powerful circumnuclear starburst ring, {and whose nuclear engine origin} is still controversial. We aim at detecting and characterizing compact radio structures both in the nuclear region and in the circumnuclear ring. We do not find any compact source in the central 200\,pc region, setting a very tight $5\sigma$ upper limit of $3.7\times10^{36}$\,erg\,s$^{-1}$ and $5.8\times10^{36}$\,erg\,s$^{-1}$, at 5.0 and 8.4\,GHz, respectively.
However, we report a clear detection at both frequencies of a compact structure in the circumnuclear ring, $190\,$pc to the north of the nucleus, whose luminosity and spectral index are compatible with a core-collapse supernova, giving support to the high star formation rate in the ring. Our result favors the pure starburst scenario, even for the nucleus of NGC\,1614, and shows the importance of radio VLBI observations when dealing with the obscured environments of dusty galaxies.

\end{abstract}


%%%%%%%%%%%%%%%%%%%%
\section{Introduction}\label{sec:intro}
%%%%%%%%%%%%%%%%%%%%

Galaxies with {infrared (IR)} luminosities higher than $10^{11} L_\odot$ (Luminous Infra-Red Galaxies; LIRGs) are known to dominate the star formation (SF) at  redshift $z\sim$1--2 \citep[e.g.,][]{magnelli09}. The study of the physics of these objects is thus essential to understand the star formation history of the Universe. Moreover, most local LIRGs contain an active galactic nuclei (AGN), which generally does not dominate their bolometric luminosity \citep{alonso-herrero12}. {Very Long Baseline Interferometry (VLBI)} observations provide a unique tool  to study in detail the nearby sources that display similar properties to those dominating the SF activity at $z\sim$1--2, i.e., local LIRGs.
Indeed, since LIRGs are heavily enshrouded in dust, VLBI observations are among the few tools that can probe deep into the innermost pc-size regions of LIRGs, thus allowing to pinpoint the precise location of the putative AGN 
as well as determining the contribution of the starburst and the AGN.  For example, \citet{perez-torres10} used contemporaneous {European VLBI Network (EVN)} observations at 1.6 and 5.0\,GHz to discover a low-luminosity AGN in Arp299 surrounded by dozens of supernovae (SNe) and supernova remnants.

NGC\,1614 is a LIRG at a distance of 64\,Mpc with an infrared luminosity of $4\times10^{11} L_\odot$, and in a late stage of a merging process. The central kpc region of NGC\,1614 hosts a prominent circumnuclear ring of star formation of $\sim600$\,pc diameter, revealed in Pa$\alpha$ \citep{alonso-herrero01}.

In \citet{herrero-illana14} we used multiwavelength observations aimed at characterizing the circumnuclear region, finding an average star formation rate (SFR) of $58\,M_\odot\,\mathrm{yr}^{-1}$ and a core-collapse SN rate of $0.4\,\mathrm{SN\,yr}^{-1}$, obtained through {spectral energy distribution (SED)} modeling. We also disentangled the radio thermal free-free and non-thermal synchrotron emission to {determine the age of} the starburst, finding an age differentiation in the ring, with the starburst in the southeast regions being younger ($\lesssim5.5$\,Myr) than that in the northwest ($\simeq8$\,Myr).

It is widely accepted that NGC\,1614 is mainly powered by a starburst. Several recent {millimeter} studies have been performed to determine the properties of the fuel of the star formation, i.e., the molecular gas, and its inflows and outflows {\citep{olsson10,imanishi13,xu15,garcia-burillo15,konig16,saito16,saito17}}. However, while the bulk of observational evidence favors a pure starburst scenario for NGC\,1614, the existence of {an intrinsically weak} AGN at its very center has not yet been completely ruled out \citep{risaliti00,wilson08,pereira-santaella15b}.

In this letter we present dual-frequency European VLBI Network (EVN) radio observations of the complete circumnuclear region aimed at obtaining a direct evidence or a tight upper limit to the luminosity of any putative AGN, as well as to detect and characterize any other compact radio structure across the circumnuclear ring.

%%%%%%%%%%%%%%%%%%%%
\section{Observations and data reduction}\label{sec:obs}
%%%%%%%%%%%%%%%%%%%%

We observed NGC\,1614 with the European VLBI Network (EVN) at 5.0\,GHz (C-band) and 8.4\,GHz (X-band), under project EH030 (PI: Herrero-Illana) in May/June 2015. The observations were contemporaneous with the aim of detecting any possible compact source and characterizing its spectral shape. We used a bandwidth of 128\,MHz for both setups and a data recording of 1024 Mbps with 2-bit sampling. We observed 3C454.3 for {fringe-fitting} and bandpass purposes, and J0427-0700 for phase calibration. In Table~\ref{tab:obs} we present a summary of the observations.

\begin{table*}
\caption{Observations summary. The codes for the participating stations are as follows (name, diameter, country): Ef: Effelsberg, 100\,m, Germany; Wb: Westerbork, 25\,m, Netherlands; O6: Onsala, 20\,m, Sweden; Mc: Medicina, 32\,m, Italy; Nt: Noto, 32\,m, Italy; Sv: Svetloe, 32\,m, Russia; Zc: Zelenchukskaya, 32\,m, Russia; Hh: Hartebeesthoek, 26\,m, South Africa; Ys: Yebes, 40\,m, Spain; Jb: Jodrell Bank, 76\,m, UK; O8: Onsala, 25\,m, Sweden; Tr: Torun, 32\,m, Poland. {The maximum recoverable scale (MRS) is obtained as $0.6\,\lambda/B_{\mathrm{min}}$, where $B_{\mathrm{min}}$ is the shortest baseline of the array.}}
\begin{tabular}{ccclccccc}
  	\hline
	Code & Band & Date & EVN participating stations & $t_\mathrm{on-source}$ &FWHM & PA & {MRS} & rms \\
   	 &  &  &  &  (min) & (mas) & (deg) & {(mas)} & ($\mu$Jy\,beam$^{-1}$)\\
	\hline
EH030A &  X  &  2015-05-29  &  Ef, Wb, O6, Mc, Nt, Sv, Zc, Hh, Ys  & 198 & $4.0\times1.9$ & $-14.6$ & 17 & 28 \\
EH030B &  C  &  2015-06-06  &  Ef, Wb, Jb, O8, Mc, Nt, Tr, Sv, Zc, Hh, Ys  & 218 & $5.2\times2.8$ & 1.1 & 28 & 30 \\
	\hline
\end{tabular}
\label{tab:obs}
\end{table*}

For data reduction, we used the NRAO Astronomical Image Processing System (AIPS). The process consisted in careful visual inspection and editing, amplitude calibration through system temperatures, and standard phase calibration.
We derived amplitude and phase gain corrections via
self-calibration on the phase-reference source, that were
later interpolated to the target source.
Imaging was also performed on AIPS, using the CLEAN algorithm {(task IMAGR)}, with a {Briggs weighting using a ROBUST parameter of 0.5}, resulting in synthesized beams of $4.7\times2.0$\,mas$^2$ ($1.61\times0.87$\,pc$^2$) and $4.0\times1.9$\,mas$^2$ ($1.24\times0.59$\,pc$^2$) and rms of 30 and 28\,$\mu$Jy\,beam$^{-1}$ for C- and X-band, respectively. No self-calibration was applied.
{We have estimated a 5\% uncertainty in the flux density calibration.}

While our main goal was to study the innermost region in search for a putative AGN, we imaged a larger field of view (4\,arcsec$^2$) so as to detect possible compact structures in all the circumnuclear ring of the galaxy.

%%%%%%%%%%%%%%%%%%%%
\section{Results and discussion}\label{sec:results}
%%%%%%%%%%%%%%%%%%%%

\begin{figure*}
\includegraphics[width=0.84\textwidth]{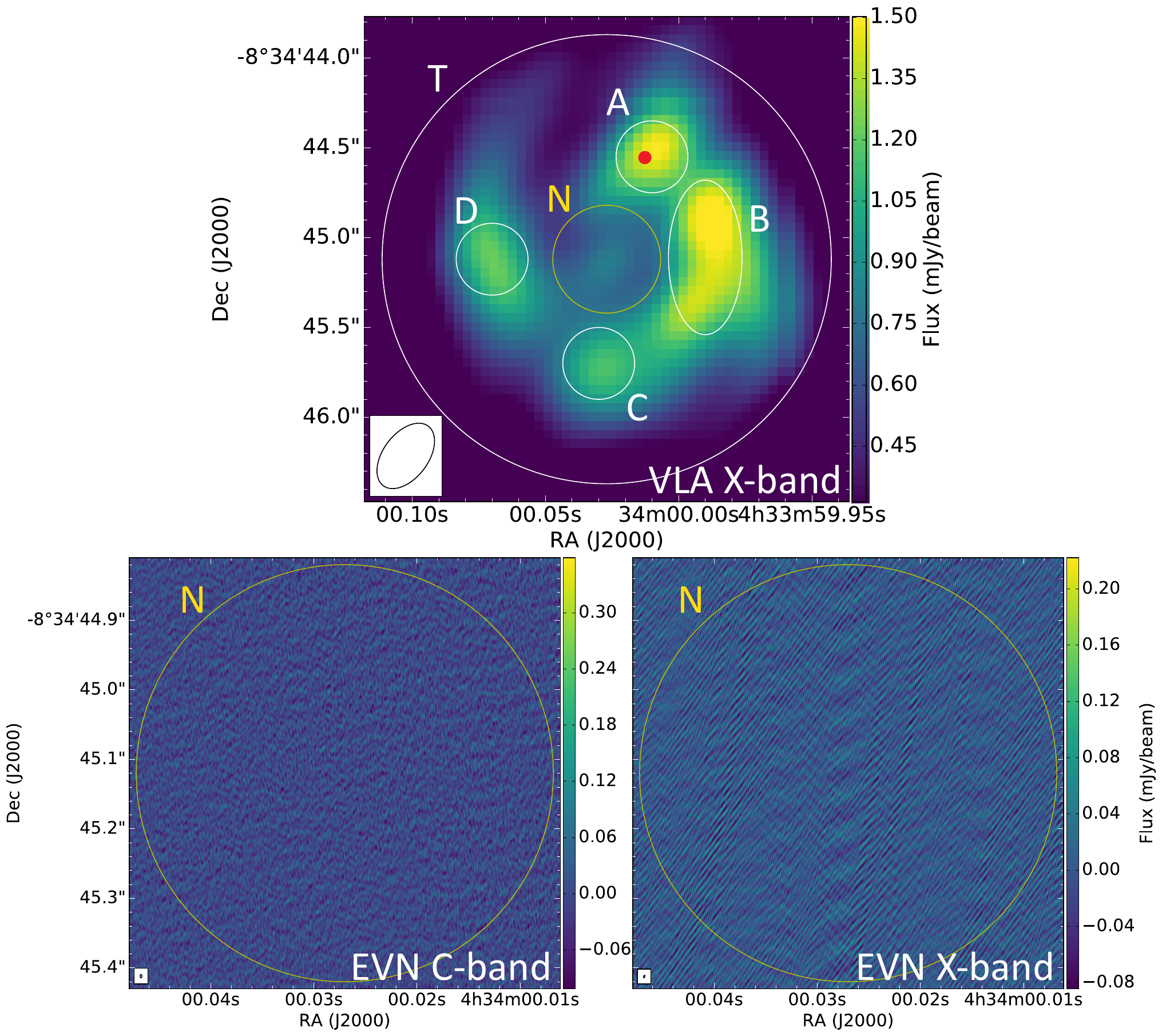}
\caption{Top panel: VLA X-band image of the central kiloparsec region of NGC\,1614, with overlaid regions as defined in \citet{herrero-illana14}. The synthesized beam size is $0.42^{''}\times 0.25^{''}$ ($130\,\mathrm{pc}\times78\,\mathrm{pc}$). The red dot indicates the position of the EVN compact source (see section~\ref{sec:nature}). Bottom panels: EVN maps at C-band (left) and X-band (right) of the nuclear region N, as labeled in the top panel. Synthesized beam sizes are shown in the bottom left corners: $5.2\,\mathrm{mas}\times 2.8\,\mathrm{mas}$ ($1.61\,\mathrm{pc}\times0.87\,\mathrm{pc}$) for C-band and $4.0\,\mathrm{mas}\times 1.9\,\mathrm{mas}$ ($1.24\,\mathrm{pc}\times0.59\,\mathrm{pc}$) for X-band. The attained rms are 30 and 28\,$\mu$Jy\,beam$^{-1}$, for C- and X-band, respectively.}
\label{fig:regionN}
\end{figure*}

In Figure~\ref{fig:regionN} we show our VLA image of the circumnuclear ring of NGC\,1614 with overlaid regions \citep[for the formal definition of the regions, see Table 2 from][]{herrero-illana14}, together with the EVN observations presented in this study of the innermost nuclear region (N). We do not find any emission in that region above $5\,\times\,$rms at any of the observed frequencies. The implications of these non-detections are developed in section~\ref{sec:agn}.

\begin{figure*}
\includegraphics[width=.93\textwidth]{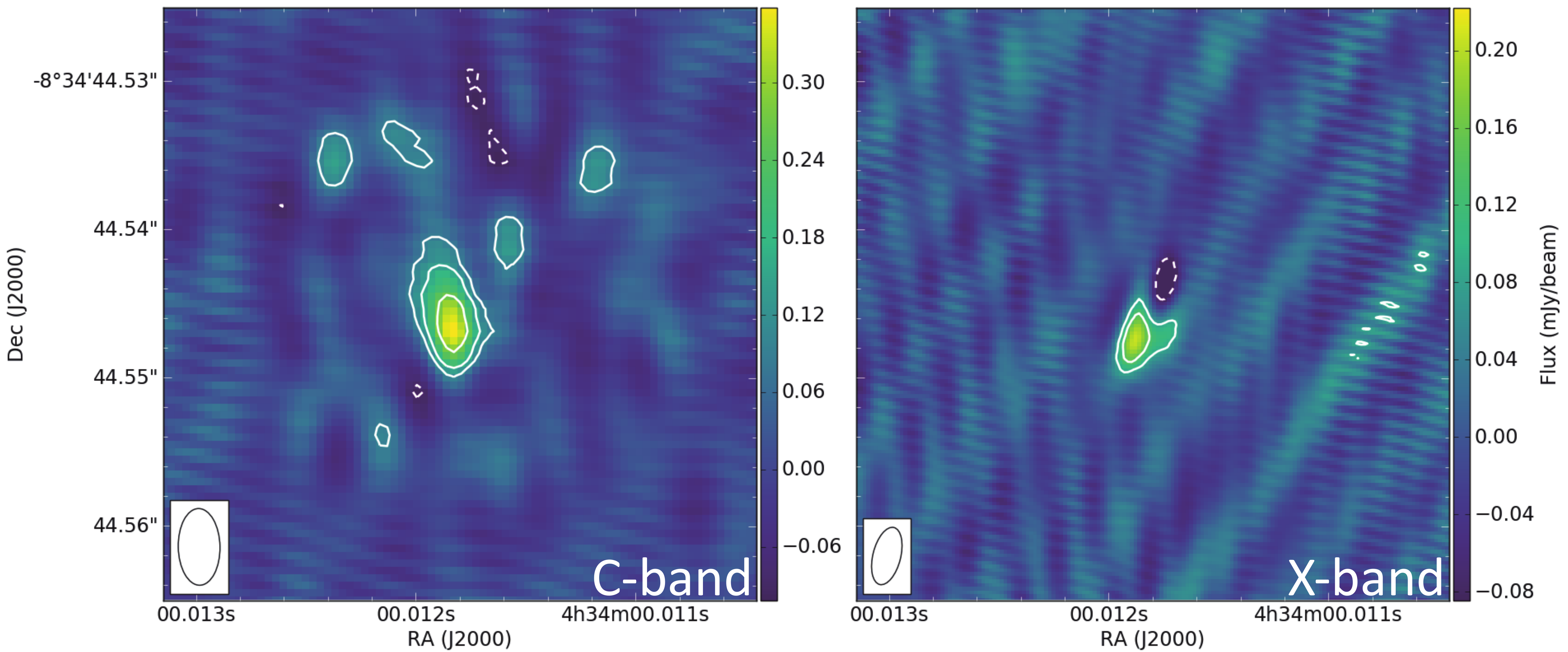}
\caption{EVN maps of the detected compact source at C-band (5.0\,GHz; left) and X-band (8.4\,GHz; right). Contours are plotted at (-3, 3, 5, 9) times the rms of each image ($30\,\mu$Jy\,beam$^{-1}$ and $28\,\mu$Jy\,beam$^{-1}$ for C- and X-band, respectively). Negative contours are dashed. The synthesized beam of each image is shown in the bottom left corners: $5.2\,\mathrm{mas}\times 2.8\,\mathrm{mas}$ ($1.61\,\mathrm{pc}\times0.87\,\mathrm{pc}$) for C-band and $4.0\,\mathrm{mas}\times 1.9\,\mathrm{mas}$ ($1.24\,\mathrm{pc}\times0.59\,\mathrm{pc}$) for X-band. The peak flux densities are $367$ and $215\,\mu\mathrm{Jy\,beam}^{-1}$ for C- and X-band, respectively.
}
\label{fig:snzoom}
\end{figure*}

There is, however, one significant detection above $5\,\times\,$rms in our imaged 4\,arcsec$^2$ region. It corresponds to a compact source in region A \citep[see][]{herrero-illana14}, which is clear at both X- and C-band. We show the corresponding images in Figure~\ref{fig:snzoom}, while the nature of this source is discussed in section~\ref{sec:nature}.

%%%%%%%%%%%%%%%%%%%%
\subsection{The elusive nuclear engine of NGC\,1614} \label{sec:agn}
%%%%%%%%%%%%%%%%%%%%

The non-detection of any nuclear compact radio structure in our deep EVN observations imposes a tight limit to the radio luminosity of a putative AGN. {\citet{olsson10} found an unresolved component ($<0.2^{\prime\prime}$) in the center of the ring at 5\,GHz using MERLIN.
However, that emission is resolved out in our higher angular resolution data, pointing to a non-compact star-formation origin, in agreement with their interpretation.}

We have established a $5\sigma$ luminosity ($\nu L_\nu$) upper limit for an AGN located in the nuclear region of NGC\,1614 of $3.7\times10^{36}$\,erg\,s$^{-1}$ in C-band and $5.8\times10^{36}$\,erg\,s$^{-1}$ in X-band, ruling out a luminous AGN, {with typical radio luminosities of $\sim10^{42}$\,erg\,s$^{-1}$ \citep[see e.g.][]{sikora07}}. {Assuming a flat spectral index between 5.0 and 1.4\,GHz, and a far-IR-to-radio correlation {\citep{condon92}} with no dispersion and the same for the AGN and for the overall emission \citep[$q=2.46$;][]{herrero-illana14}, we derive an AGN contribution of $\lesssim0.1\%$ relative to the total IR luminosity.}

However, the possibility of a low luminosity AGN (LLAGN) still remains. While most LLAGN have detected radio cores with luminosities above our threshold \citep{nagar02}, there are several compact detections of very nearby galaxies in the low luminosity end of LLAGN that have significantly lower luminosities \citep[e.g. NGC\,4051, with $\nu L_\nu=10^{35}$\,erg\,s$^{-1}$ at 5\,GHz;][]{giroletti09}.

\citet{pereira-santaella15b} found that the mid-IR properties of the 150\,pc nuclear region of NGC\,1614 are at odds with those from the starburst ring. Interestingly, the $11.3\,\mu$m PAH equivalent width and the mid-IR continuum are similar to those found in Seyfert galaxies, proposing an intrinsically X-ray weak AGN as a plausible scenario. The relation between X-ray and radio emission can be used to characterize AGN \citep{terashima03}. Using our 5\,GHz luminosity upper limit and assuming that most of the hard X-ray emission \citep[$1.4\times10^{41}$\,erg\,s$^{-1}$ within a region of $\simeq1.5\,$kpc;][]{pereira-santaella15b} is due to the AGN, NGC\,1614 would be compatible with a LLAGN in \citet{terashima03} diagnostic plots. However, according to the X-ray and radio 6\,cm relation by \citet{panessa07}, the expected 6\,cm luminosity would be $\left(1.1^{+4.0}_{-0.8}\right)\times10^{37}$\,erg\,s$^{-1}$, possibly but unlikely missed by our observations. Anyhow, since most of the hard X-ray luminosity can be explained with star formation \citep{pereira-santaella11}, we find more plausible that NGC\,1614 is being essentially powered by stellar activity and/or supernova shocks.

%%%%%%%%%%%%%%%%%%%%
\subsection{The nature of the off-nuclear compact source}\label{sec:nature}
%%%%%%%%%%%%%%%%%%%%

The only significant component detected in our large field of view is located at $04^\mathrm{h}34^\mathrm{m}00.^\mathrm{s}0118 \,-08^\circ34^\prime44.^{\prime\prime}547$, which lays very close to the center of region A (see top panel in Fig.~\ref{fig:regionN}), and is located at $0.62$\,arcsec from the center of the galaxy, which corresponds to a projected distance of 192\,pc. We show a zoomed map of this area in Figure~\ref{fig:snzoom}. The peak flux density, $S_\nu$, of this compact {unresolved} source is $367\,\mu\mathrm{Jy\,beam}^{-1}$ at C-band , and $215\,\mu\mathrm{Jy\,beam}^{-1}$ at X-band. These flux densities correspond to $12\sigma$ and $8\sigma$ detections for C- (5.0\,GHz) and X-band (8.4\,GHz), respectively. We note however, that the X-band data have some sort of phase calibration problem, as can be seen in the figure, that we were not able to remove completely. The derived two-point spectral index, $\alpha$ (where $S_\nu\propto\nu^\alpha$) is $-1.0\pm0.3$. The quoted flux densities, {including the systematic uncertainty,} correspond to $(9.0\pm0.9)\times10^{36}$\,erg\,s$^{-1}$ (C-band) and $(8.9\pm1.2)\times10^{37}$\,erg\,s$^{-1}$ (X-band).

The three most plausible explanations for the nature of this source are: (i) an off-nuclear AGN, (ii) a compact super star cluster (SSC), or (iii) a supernova or supernova remnant. The three possibilities are discussed here:

\begin{enumerate}
\item An AGN is not necessarily located in the innermost nuclear region of a galaxy. On the one hand, NGC\,1614 is a late stage merger, with a secondary component that could host an AGN itself. However, the secondary point source revealed in near-IR images \citep[0.90\,arcsec to the north-east of the main nucleus;][]{alonso-herrero01, dametto14}, and not detected in our EVN observations, is  0.89\,arcsec to the east of our detected compact structure.
On the other hand, although in some cases the AGN is not located at the center of the circumnuclear disk \citep[e.g. NGC\,1068;][]{garcia-burillo14}, off-nuclear AGN are often associated to recoiling black holes \citep{volonteri08} in dual AGN systems, which is unlikely the case. 
Finally, AGN typically exhibit flat radio spectral indices ($\alpha\sim0$), unlike our derived index of $\alpha=-1.0\pm0.3$.

\item We can also exclude a SSC nature based on the argument followed by \citet{perez-torres09b} regarding the bright temperature, $T_B$ of the source, which is given by:

\begin{equation}
T_B=\frac{2c^2B_\nu}{k\nu^2},
\end{equation}
where $B_\nu$ is the intensity measured in erg\,s$^{-1}$\,Hz$^{-1}$\,str$^{-1}$. This depends of the deconvolved size of the source, which we estimate to be $6.1\times3.2$\,mas$^2$ for our C-band image based on a simple Gaussian fit. We find $T_B=1.25\times10^6$\,K, an unfeasible value for the thermal emission originated in a SSC. Additionally, the maximum number of ionizing O stars in such a cluster would produce a thermal luminosity that is still two orders of magnitudes below the observed luminosity \citep{herrero-illana12a}.

\item As found in \citet{herrero-illana14} based on the thermal vs non-thermal emission relation, regions A and B (see Fig.~\ref{fig:regionN}, top) are the most prone to harbor supernovae. In particular, region A, where this compact source is located, is a young ($\sim8$\,Myr), synchrotron-dominated region, likely powered by supernovae. This, together with the large SN rate \citep[$0.4\,\mathrm{SN\,yr}^{-1}$;][]{herrero-illana14}, {the short gas depletion time of the ring \citep{saito16}}, our derived synchrotron spectral index $\alpha=-1.0\pm0.3$ and the obtained luminosity, supports the idea that the source has a core-collapse supernova nature. The estimated starburst age of $\sim8$\,Myr constrains the initial mass of the progenitor star in the $\simeq20-30\,M_\odot$ range.
The detection of supernovae and supernova remnants in enshrouded environments by means of VLBI observations is becoming common with the enhancement in sensitivity of radio interferometers \citep[see e.g.][]{neff04, perez-torres09b,ulvestad09,batejat11,bondi12,romero-canizales12b,romero-canizales17}. While the nature of the core-collapse supernova cannot be unambiguously determined without constraining its light curve through a monitoring of the source, the steep spectral index indicates that its radio brightness, $L_{5\,\mathrm{GHz}}=(1.8\pm0.2)\times10^{27}$\,erg\,s$^{-1}$\,Hz$^{-1}$, is already diminishing. Furthermore, its high luminosity likely rules out a type IIP supernova \citep{chevalier06}, which is the most abundant type in the local Universe \citep[$\simeq59\%$ of all SNe;][]{smartt09a}.

The number of SNe found in LIRGs by means of VLBI observations, considering the very few studies available, is uneven. An interesting case is Arp~299, as its IR luminosity ($6.7\times10^{11}L_\odot$) and distance (45\,Mpc) make it comparable to NGC\,1614. From the dozens of compact sources detected in Arp~299 \citep{perez-torres09b,bondi12}, only eight are above our 5$\sigma$ detection threshold ($7.4\times10^{26}\,$erg\,s$^{-1}$\,Hz$^{-1}$) and have a SN origin. Furthermore, the supernova rate obtained through consistent SED multi-wavelength modeling is higher for Arp~299 \citep[$\simeq0.8$\,SN\,yr$^{-1}$;][]{mattila12} than for NGC\,1614 \citep[$\simeq0.4$\,SN\,yr$^{-1}$;][]{herrero-illana14}. Therefore, we would expect four SN detections in our observations. While the low number statistics may explain this discrepancy, differences in the star formation history \citep{pereira-santaella15a}, in the environment \citep{klessen07}, or even in the initial mass function \citep{perez-torres09b} cannot be ruled out. In this sense, we are currently analyzing VLBI data on a sample of 10 LIRGs (P\'erez-Torres et al. in prep.) for which we have studied their starburst and AGN properties, as well as their star formation histories using different diagnostics \citep{herrero-illana17}.

\end{enumerate}

The scenario of a powerful star formation ring surrounding a non-active nucleus can be explained by a balance between the in-falling and out-falling gas or by the molecular gas exhaustion in the ring due to the intense star formation \citep{perez-torres10, saito17}, preventing the supermassive black hole from being fed.
This is in agreement with the SF-powered molecular outflow found in NGC\,1614 by \citet{garcia-burillo15}.
It is also possible that a massive inflow of gas caused by the final phase of the merging event has not yet been fully funneled to the central region, which will eventually cause and enhancement in star formation, and an active phase of the AGN \citep{konig16}.
In any case, our finding supports the model where mechanical heating from SN shocks and superwinds can heat the surrounding medium, inhibiting the accretion of material towards the central AGN and powering the warm gas in the starburst ring of NCG1614 \citep{saito17}.

%%%%%%%%%%%%%%%%%%%%
\section{Summary}\label{sec:summary}
%%%%%%%%%%%%%%%%%%%%

We have obtained radio VLBI observations of the central kiloparsec region of NGC\,1614 with the aim of detecting and characterizing a putative AGN and any other compact structure in the circumnuclear ring. 

We do not detect any compact structure in the innermost 200 parsec region. We have established a tight $5\sigma$ upper limit luminosity for any putative AGN in this region of $3.7\times10^{36}$\,erg\,s$^{-1}$ and $5.8\times10^{36}$\,erg\,s$^{-1}$ at 5.0 and 8.4\,GHz, respectively, ruling out any significant contribution in radio wavelengths. {Furthermore, we estimate an AGN contribution of $\lesssim 0.1\%$ in the IR.} There is also no hint of any compact radio source in the proposed secondary nucleus of the system.

We detect a compact source in the circumnuclear ring of NGC\,1614 located $\simeq0.6$\,arcsec ($\simeq190\,$pc) to the north of the nucleus. Its luminosity ($9.0\times10^{36}$\,erg\,s$^{-1}$ and $8.9\times10^{36}$\,erg\,s$^{-1}$ at C- and X-band, respectively) and spectral shape ($\alpha\simeq-1$) are indicative of a supernova origin, possibly a type IIn or Ic. Its location in one of the knots of star formation of the circumnuclear ring points towards the same direction. This supports models using mechanical heating to warm the observed molecular gas in the ring.

In summary, while we cannot rule out completely the existence of an (extremely radio faint) AGN in the central $\sim200$\,pc of the nucleus in NGC\,1614, its contribution to the radio {and IR emissions} of the galaxy is negligible. This result, together with the compact and diffuse radio emission in its circumnucleear $\sim$kpc region, indicates a virtually pure starburst nature for NGC\,1614.

\section*{Acknowledgements}
RHI, AA, and MAPT acknowledge  support from the Spanish MINECO through grant AYA2015-63939-C2-1-P, co-funded with FEDER funds.
AAH acknowledges support through grant  AYA2015-64346-C2-1-P, co-funded with FEDER funds.
MPS acknowledges support from STFC through grant ST/N000919/1, from the John Fell Oxford University Press (OUP) Research Fund, and from the University of Oxford.
The European VLBI Network is a joint facility of independent European, African, Asian, and North American radio astronomy institutes. Scientific results from data presented in this publication are derived from the following EVN project code(s): EH030.

\bibliographystyle{mn2e}
\small
\bibliography{/Users/rherrero/Dropbox/masterbibdesk}

% Don't change these lines
%\bsp	% typesetting comment
\label{lastpage}
\end{document}